\documentclass[12pt]{article}

\pdfoutput=1

\usepackage{ifpdf}
\ifpdf
\usepackage{graphicx,color}
\usepackage{hyperref}
\else
\usepackage[dvipdfmx]{graphicx,color}
\usepackage[dvipdfmx]{hyperref}
\fi
\usepackage{amssymb,amsfonts,amsmath,cancel,cite,multirow}
\usepackage[capitalise]{cleveref}
\usepackage{slashed}

\newcommand{\eqs}[1]{\begin{equation}\begin{split} #1 \end{split}\end{equation}}

\begin{document}

\begin{titlepage}

\begin{flushright}
	CTPU-PTC-19-24
\end{flushright}

\vskip 1.35cm
\begin{center}

{\large
\textbf{
Lessons from $T^{\mu}_{~ \mu}$ on inflation models: \\
two-scalar theory and Yukawa theory}}
\vskip 1.2cm

Ayuki Kamada$^{a}$
and Takumi Kuwahara$^{a}$

\vskip 0.4cm

\textit{$^a$
Center for Theoretical Physics of the Universe,
Institute for Basic Science (IBS), Daejeon 34126, Korea
}

\vskip 1.5cm

\begin{abstract}
We demonstrate two properties of the trace of the energy-momentum tensor $T^{\mu}_{~ \mu}$ in the flat spacetime.
One is the decoupling of heavy degrees of freedom; i.e., heavy degrees of freedom leave no effect for low-energy $T^{\mu}_{~ \mu}$-inserted amplitudes.
This is intuitively apparent from the effective field theory point of view, but one has to take into account the so-called trace anomaly to explicitly demonstrate the decoupling.
As a result, for example, in the $R^{2}$ inflation model, scalaron decay is insensitive to heavy degrees of freedom when a matter sector {\it minimally} couples to gravity (up to a non-minimal coupling of a matter scalar field other than the scalaron).
The other property is a quantum contribution to a non-minimal coupling of a scalar field.
The non-minimal coupling disappears from the action in the flat spacetime, but leaves the so-called improvement term in $T^{\mu}_{~ \mu}$.
We study the renormalization group equation of the non-minimal coupling to discuss its quantum-induced value and implications for inflation dynamics.
We work it out in the two-scalar theory and Yukawa theory.
\end{abstract}

\end{center}
\end{titlepage}

\section{Introduction}

The energy-momentum tensor $T_{\mu \nu}$ is an important object in quantum field theory~\cite{Callan:1970ze, Coleman:1970je}.
It provides generators of spacetime symmetry (Poincar\'e symmetry in the flat spacetime).
It is conserved, $\nabla_{\mu} T^{\mu \nu} = 0$ ($\nabla_{\mu}$: diffeomorphism covariant derivative), and finite up to an improvement term.
Its trace $T^{\mu}_{~ \mu}$ is the divergence of the dilatation current and vanishes when matter respects scale invariance.
Moreover, it determines the coupling of matter to gravity.

The aim of this article is to highlight mainly two properties of $T^{\mu}_{~ \mu}$ in the flat spacetime.
One is the decoupling of heavy degrees of freedom.
Low-energy $T^{\mu}_{~ \mu}$-inserted amplitudes are robustly calculated in terms of effective field theory.
The other is the renormalization of an improvement term of $T^{\mu}_{~ \mu}$.
In this article, we define the energy-momentum tensor by a functional derivative of the matter action with respect to a metric $g_{\mu \nu}$.
The matter action is assumed to be ``minimally'' coupled to gravity up to a non-minimal coupling of the scalar field to gravity, $\xi R \phi^{2}$ ($R$: Ricci scalar, $\phi$: scalar field).
The improvement term of $T^{\mu}_{~ \mu}$, $2 \eta \partial^{2} \phi^{2}$, originates from the non-minimal coupling.
Here $\xi = \xi_{c} + \eta / (d - 1)$ with the conformal coupling $\xi_{c} = (d - 2) / (4 (d - 1))$ in $d$ dimensions.
We work it out in the two-scalar theory and Yukawa theory.

These properties are relevant when one considers inflation models.
The decoupling of heavy degrees of freedom is important when one considers the reheating of $R^{2}$ inflation (or generically $f(R)$ inflation)~\cite{Starobinsky:1980te, Barrow:1983rx, Whitt:1984pd, Vilenkin:1985md, Mijic:1986iv, Barrow:1988xh}.%
\footnote{
Our discussion is also applicable to $f(\sigma) R$ inflation ($\sigma$: inflaton)~\cite{Spokoiny:1984bd, Accetta:1985du, Lucchin:1985ip, La:1989za, Futamase:1987ua, Salopek:1988qh, Fakir:1990eg}, when $\sigma$ does not couple to matter in the Jordan frame for some reason.
}
In this model, a Weyl degree of freedom of the metric in the Jordan frame, called the scalaron, is identified as the inflaton.
Through a Weyl transformation one can move to the {\it scalaron frame}, where gravity has the Einstein-Hilbert action (up to a non-minimal coupling of a matter scalar field to $R$) and the scalaron has a canonical kinetic term and couples to matter through $T^{\mu}_{~ \mu}$~\cite{Whitt:1984pd, Jakubiec:1988ef, Faulkner:2006ub}.
Scalaron (inflaton) decay is determined by $T^{\mu}_{~ \mu}$-inserted amplitudes.

On the other hand, the decoupling of heavy degrees of freedom from low-energy $T^{\mu}_{~ \mu}$-inserted amplitudes is not obvious at first sight.
This is because $T^{\mu}_{~ \mu}$ consists of mass terms (classical breaking of scale invariance).
Some loop diagrams with the insertion of a mass term do not vanish in the heavy mass limit, leaving non-decoupling contributions at low energy.
A key is to take into account the quantum breaking of scale invariance, known as the trace anomaly~\cite{Callan:1970ze, Coleman:1970je, Freedman:1974gs, Freedman:1974ze, Collins:1976vm, Nielsen:1977sy, Adler:1976zt, Collins:1976yq, Brown:1979pq, Brown:1980qq, Hathrell:1981zb, Hathrell:1981gz}.%
\footnote{
The trace of the energy-momentum tensor and trace anomaly are often not distinguished.
In this paper we use the former to refer to the whole (classical + quantum) contribution, while we use the latter to refer to only a quantum contribution.
}
The cancellation between the contributions from the classical breaking and from the quantum breaking is explicitly demonstrated for the gauge trace anomaly~\cite{Kamada:2019pmx}.%
\footnote{
Ref.~\cite{Kamada:2019pmx} works in $R^{2}$ inflation to be concrete, while it is applicable to a broad class of scalar-tensor gravity as discussed above.
See Ref.~\cite{Choi:2019osi} for $f(\sigma) R$ inflation.
}

The importance of the renormalization of the improvement term is evident from the following view point.
Measurements of cosmic microwave background (CMB) anisotropies~\cite{Hinshaw:2012aka, Akrami:2018odb} disfavor chaotic inflation with a simple power-law potential~\cite{Linde:1983gd} due to the predicted large tensor-to-scalar ratio.
The situation gets improved simply with a small non-minimal coupling of $\xi \sim - 10^{-3}$~\cite{Linde:2011nh, Boubekeur:2015xza, Shokri:2019rfi}.%
\footnote{
The minus sign originates from our convention following Refs.~\cite{Kolb:1990vq, Hathrell:1981zb}: the metric signature is $(+ , - , - , -)$; the Einstein-Hilbert action with a free singlet scalar is
\begin{eqnarray}
S_{\text{E-H}} = - \frac{M_{\rm pl}^{2}}{2} \int d^{4} x \sqrt{- g} R + \int d^{4} x \sqrt{- g} \left( \frac{1}{2} g^{\mu \nu} \nabla_{\mu} \phi \nabla_{\nu} \phi + \frac{1}{2} \xi R \phi^{2} \right)  \,,
\end{eqnarray}
with the reduced Planck mass $M_{\rm pl}$;
and the four-dimensional conformal coupling is $\xi_{c} = + 1/6$.
}
While one can take any value of $\xi$ at the classical level, it is quite intriguing if such a small $\xi$ appears at the quantum level.
The quantum-induced value of $\eta$ is studied in the $\lambda \phi^{4}$ theory~\cite{Collins:1976vm, Brown:1980qq, Hathrell:1981zb}: $\Delta \eta = - \lambda^{3} / (864 (4\pi)^{6})$ at the leading (three-loop) order.
It appears from the renormalization of trace-anomaly terms (i.e., composite operators~\cite{Zimmermann:1969jj, Lowenstein:1971jk, Collins:1974da, Breitenlohner:1977hr}) and is related with an inhomogeneous term of the $\beta$ function of $\eta$.
Unless the leading value is at the one-loop order, it is hard to imagine that $\xi \sim - 10^{-3}$ originates from a quantum contribution.
We find that $\Delta \eta$ appears at the one-loop order in the two-scalar theory and Yukawa theory, as a threshold correction when additional degrees of freedom are heavy and decouple from the low-energy dynamics.
Nevertheless, its sign is positive and thus opposite from that required for chaotic inflation.
Meanwhile, an inhomogeneous term of the $\beta$ function of $\eta$ does not appear at the one-loop level.

This article is organized as follows.
In the next section, we evaluate $T^{\mu}_{~ \mu}$-inserted diagrams.
We demonstrate how the contribution from trace-anomaly terms (quantum breaking of scale invariance) cancels with that from mass terms of heavy degrees of freedom (classical breaking of scale invariance).
We find that heavy degrees of freedom leave a one-loop threshold correction to $\eta$, which is regarded as the quantum-induced value of $\eta$.
In \cref{sec:naturaleta} we study the renormalization group equation (RGE) of $\eta$ to discuss the quantum-induced value of $\eta$ when additional degrees of freedom do not decouple.
\cref{sec:concl} is devoted to a summary and further remarks.
Throughout this article, we adopt the modified minimal subtraction ($\overline{\rm MS}$) scheme~\cite{Ashmore:1972uj, Bollini:1972ui, tHooft:1972tcz} with the spacetime dimension of $d = 4 - \epsilon$ and the (modified) renormalization scale $\mu$ ($\tilde \mu$).
We summarize related one-loop calculations in \cref{sec:oneloop}.

\section{Decoupling of heavy degrees of freedom \label{sec:decoupling}}
We assume that the matter sector is minimally coupled to gravity, while maintaining renormalizability%
\footnote{This does not mean the matter sector consists solely of a finite number of renormalizable terms.
Non-renormalizable terms are allowed when an infinite number of non-renormalizable terms are introduced for renormalization in the usual sense of effective field theory.} up to graviton loops that are suppressed by $1 / M_{\rm pl}^{2}$:
\eqs{
S_{\rm mat} \left[ \{ \phi_{0 i} \}, g_{\mu \nu}; \{ \lambda_{0 a} \} \right] \,,
}
where $\{ \phi_{0 i} \}$ and $\{ \lambda_{0 a} \}$ collectively denote bare matter fields and parameters, respectively.
In particular we require the renormalizability of the energy-momentum tensor that is defined as a linear response of the matter action to the metric:
\eqs{
\label{eq:Tmunu}
T^{\mu \nu} = - \frac{2}{\sqrt{-g}} \frac{\delta S_{\rm mat} \left[ \{ \phi_{0 i} \}, g_{\mu \nu}; \{ \lambda_{0 a} \} \right]}{\delta g_{\mu \nu}} \,.
}

In the following, we take into account gravity only to derive $T^{\mu \nu}$.
We consider a non-minimal coupling $\xi$ as a part of $S_{\rm mat}$.
We evaluate the trace of the energy-momentum tensor $T^{\mu}_{~ \mu}$ in the flat spacetime.
We remark that in the flat spacetime, $\xi$ appears only in $T^{\mu}_{~ \mu}$ as an improvement term.
$\xi$ does not change the usual multiplicative renormalization of all the fields and parameters.
Thus, $T^{\mu}_{~ \mu}$ in the flat spacetime, which consists of the renormalized fields and parameters, is almost pre-determined.
The single exception is $\xi$, which is determined by the renormalization of $T^{\mu}_{~ \mu}$ itself.

In the $\overline{\rm MS}$ scheme, we first calculate $d$-dimensional $T^{\mu \nu}$, take the trace $T^{\mu}_{~ \mu}$, and then take the limit of $\epsilon \to 0$.
As stressed in Ref.~\cite{Kamada:2019pmx}, a key point is that $T^{\mu}_{~ \mu}$ contains terms proportional to $\epsilon$.
These terms vanish in the limit of $\epsilon \to 0$ at the classical level, but not at the quantum level due to the renormalization of composite operators.
This is the origin of the trace anomaly.
As we will see in explicit examples below, the trace anomaly plays an important role in the decoupling of heavy degrees of freedom.

In the following, for diagrammatic convenience, we introduce a ``scalaron'' $\sigma$ that couples to $T^{\mu}_{~ \mu}$ as
\eqs{
{\cal L}_{\sigma \text{-mat}} = \sigma T^{\mu}_{~ \mu}  \,.
}
With this coupling, we calculate the scalaron decay amplitude into light scalars $\phi$: ${\cal M} (\sigma \to \phi \phi)$.
From the Lehmann-Symanzik-Zimmermann (LSZ) reduction formula~\cite{Lehmann:1954rq}, ${\cal M}$ is given by the amputated amplitude times the product of the square-root residues of the mass pole of light degrees of freedom, $\left( Z^{\rm pole} \right)^{1/2}$'s.
In $R^{2}$ inflation, the scalaron $\sigma$ couples to $T^{\mu \nu}$ as
\eqs{
{\cal L}_{\sigma \text{-mat}} =\frac{1}{\sqrt{6}} \frac{\sigma}{M_{\rm pl}} T^{\mu}_{~ \mu} \,.
}
Thus, the corresponding invariant amplitude of scalaron decay is given by
\eqs{
\label{eq:dec}
{\cal M}_{\rm dec} = \frac{1}{\sqrt{6}} \frac{1}{M_{\rm pl}} {\cal M} \,.
}

We remark that although we calculate the scalaron decay amplitude for diagrammatic convenience, our results are not limited within the inflation model with the scalaron.
Through this decay amplitude, we study the properties of $T^{\mu}_{~ \mu}$ such as the decoupling of heavy degrees of freedom and the renormalization of a non-minimal coupling of $\phi$.
One can regard $\phi$ as the inflaton as we will do in the next section.

\subsection{Two-scalar theory}
Let us consider the following action with two real scalar fields, $\phi$ and $\psi$:
\eqs{
S_{\rm mat} =& \int d^{d} x \sqrt{- g} \left( \frac{1}{2} g^{\mu \nu} \nabla_{\mu} \phi_{0} \nabla_{\nu} \phi_{0} + \frac{1}{2} \xi_{\phi 0} R \phi_{0}^{2} + \frac{1}{2} g^{\mu \nu} \nabla_{\mu} \psi_{0} \nabla_{\nu} \psi_{0} + \frac{1}{2} \xi_{\psi 0} R \psi_{0}^{2} \right. \\
& \left. - \frac{1}{2} M^{2}_{0} \phi_{0}^{2} - \frac{1}{2} m^{2}_{0} \psi_{0}^{2}  - \frac{1}{4!} \lambda_{\phi 0} \phi_{0}^{4} - \frac{1}{4!} \lambda_{\psi 0} \psi_{0}^{4} - \frac{1}{4} \chi_{0} \phi_{0}^{2} \psi_{0}^{2}  \right) \,.
}
Parameters are scalar masses, $M$ and $m$, self-couplings, $\lambda_{s}$ ($s = \phi, \psi$), and quartic coupling $\chi$.
We summarize the multiplicative renormalization of fields and parameters and its one-loop expressions in \cref{sec:two-scalar}.
The $d$-dimensional flat-spacetime energy-momentum tensor is given by
\eqs{
T_{\mu \nu} =& \partial_{\mu} \phi_{0} \partial_{\nu} \phi_{0}
+ \partial_{\mu} \psi_{0} \partial_{\nu} \psi_{0}
- \left( \frac{d - 2}{4 (d - 1)} + \frac{\eta_{\phi 0}}{d - 1} \right) (\partial_{\mu} \partial_{\nu} - g_{\mu \nu} \partial^{2}) \phi_{0}^{2} \\
& - \left( \frac{d - 2}{4 (d - 1)} + \frac{\eta_{\psi 0}}{d - 1} \right) (\partial_{\mu} \partial_{\nu} - g_{\mu \nu} \partial^{2}) \psi_{0}^{2}
 - g_{\mu \nu} {\cal L} \,,
}
where the flat-spacetime Lagrangian density ${\cal L}$ is given by \cref{eq:two-scalar-L}.
Here we use $\xi = (d - 2) / (4 (d - 1)) + \eta / (d - 1)$.
Taking the trace, one finds
\eqs{
T^{\mu}_{~ \mu} = &\eta_{\phi 0} \partial^{2} \phi_{0}^{2} + \eta_{\psi 0} \partial^{2} \psi_{0}^{2} + M_{0}^{2} \phi_{0}^{2} + m_{0}^{2} \psi_{0}^{2} \\
& + \epsilon \left( \frac{1}{4!} \lambda_{\phi 0} \phi_{0}^{4} + \frac{1}{4!} \lambda_{\psi 0} \psi_{0}^{4}
+ \frac{1}{4} \chi_{0} \phi_{0}^{2} \psi_{0}^{2} \right) + {\rm (e.o.m.)} \,,
}
where the last term is proportional to the equation of motion:
\eqs{
{\rm (e.o.m.)} =&
\left( 1 - \frac{\epsilon}{2} \right) \phi_{0} \left[ \partial^{2} \phi_{0} + M_{0}^{2} \phi_{0} + \frac{4}{4!} \lambda_{\phi 0} \phi_{0}^{3} + \frac{2}{4} \chi_{0} \phi_{0} \psi_{0}^{2} \right] \\
& + \left( 1 - \frac{\epsilon}{2} \right) \psi_{0} \left[ \partial^{2} \psi_{0} + m_{0}^{2} \psi_{0} + \frac{4}{4!} \lambda_{\psi 0} \psi_{0}^{3} + \frac{2}{4} \chi_{0} \phi_{0}^{2} \psi_{0} \right] \,.
}

We consider the decay of the scalaron $\sigma$ into two light scalars $\phi$ at the one-loop level.
We assume that $\phi$ is much lighter than $\psi$, $M_{\rm phys} \ll m_{\rm phys}$ (pole mass), and the self-couplings, $\lambda_{\phi}$ and $\lambda_{\psi}$, are negligible.
The leading contributions originate from
\eqs{
\label{eq:two-scalar-Tmumu}
T^{\mu}_{~ \mu} \supset& \eta_{\phi} \partial^{2} \phi^{2} + \eta_{\psi} \partial^{2} \psi^{2} + M^{2} \phi^{2} + m^{2} \psi^{2} + \frac{1}{4} \epsilon {\tilde \mu}^{\epsilon} \chi \phi^{2} \psi^{2} \\
& + (Z_{\eta \phi} - 1) \eta_{\phi} \partial^{2} \phi^{2} + (Z_{M^{2}} - 1) M^{2} \phi^{2} \,.
}
Here we use the renormalized fields and parameters in \cref{eq:two-scalar-field-renom,eq:two-scalar-param-renom,eq:two-scalar-eta-renom}.
The one-loop decay amplitude is given by
\eqs{
i {\cal M} (\sigma (p) \to \phi(q) \phi(k)) = i {\cal M}^{\rm tree} + i {\cal M}^{\rm loop} + i {\cal M}^{\rm c.t.} \,,
}
where $p$, $q$, and $k$ are external momenta.
The counterterm contribution from the insertion of $T^{\mu}_{~ \mu} \supset (Z_{\eta \phi} - 1) \eta_{\phi} \partial^{2} \phi^{2} + (Z_{M^{2}} - 1) M^{2} \phi^{2}$ is given by
\eqs{
\label{eq:two-scalar-ct}
i {\cal M}^{\rm c.t.} = 2 i (Z_{M^{2}} - 1) M^{2} - 2 i (Z_{\eta \phi} - 1) \eta_{\phi} p^{2} = 2 i \frac{\chi}{16 \pi^{2}} m^{2} \frac{1}{\epsilon} - 2 i (Z_{\eta \phi} - 1) \eta_{\phi} p^{2} \,.
}
In the second equality, we use $Z_{M^{2}}$, which is determined to absorb the $1/\epsilon$ pole in the one-loop self-energy of $\phi$ and is given by \cref{eq:two-scalar-Zphi}.
While $Z_{M^{2}}$ is pre-determined, $Z_{\eta \phi}$ is the counterterm to cancel the $1/\epsilon$ pole that appears in ${\cal M}^{\rm loop}$.

The tree-level contribution from the insertion of $T^{\mu}_{~ \mu} \supset \eta_{\phi} \partial^{2} \phi^{2} + M^{2} \phi^{2}$ is given by
\eqs{
i {\cal M}^{\rm tree} &= 2 i ( M^{2} - \eta_{\phi} p^{2} ) \\
& = 2 i ( M_{\rm phys}^{2} - \eta_{\phi} p^{2} ) - i \frac{\chi}{16 \pi^{2}} m^{2} \left[ \ln \left( \frac{m^{2}}{\mu^{2}} \right) - 1 \right] \,.
}
We do not include $Z^{\rm pole}_{\phi}$ since the wave function is not renormalized at this order [see \cref{eq:two-scalar-Zphi}].
In the second line, we use the relation between the pole and renormalized masses squared of the scalar field in \cref{eq:two-scalar-delM}.
The $m^{2}$ term in ${\cal M}^{\rm tree}$ diverges as we take the heavy limit of $\psi$.
This originates from the fact that the scalar mass squared is sensitive to ultraviolet physics and one needs fine-tuning to realize $M_{\rm phys} \ll m_{\rm phys}$.
This $m^{2}$ term should be canceled by other contributions so that the amplitude is insensitive to ultraviolet physics.
We will see it shortly below.

\begin{figure}
	\centering
	\includegraphics[width=0.75\linewidth]{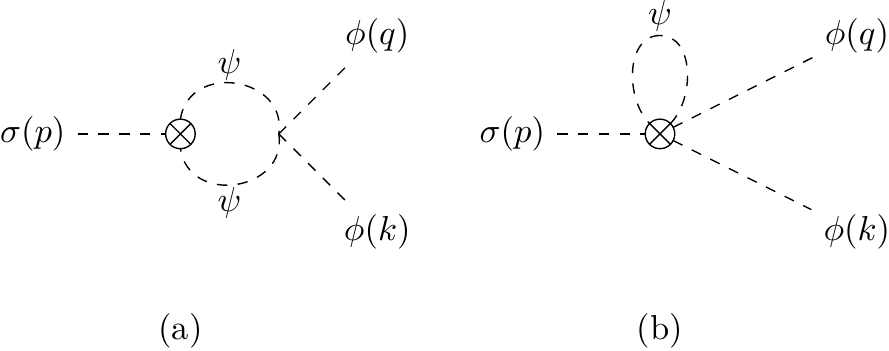}
	\caption{One-loop diagrams for scalaron decay ($\sigma \to \phi \phi$).
	Crossed dots denote the insertion of the energy-momentum tensor, $\eta_{\psi} \partial^{2} \psi^{2} + m^{2} \psi^{2}$ (a) and $(1/4) \epsilon {\tilde \mu}^{\epsilon} \chi \phi^{2} \psi^{2}$ (b).}
	\label{fig:two-scalar-loop}
\end{figure}

\cref{fig:two-scalar-loop} shows the one-loop diagrams contributing to ${\cal M}^{\rm loop}$.
We provide computational details in \cref{sec:two-scalar}.
\cref{fig:two-scalar-loop} (a) from the insertion of $T^{\mu}_{~ \mu} \supset  \eta_{\psi} \partial^{2} \psi^{2} + m^{2} \psi^{2}$ gives
\eqs{
i {\cal M}^{\rm loop}_{1}
& = - i \frac{\chi}{16 \pi^{2}} ( m^{2} - \eta_{\psi} p^{2} ) \left[ \frac{2}{\epsilon} - \ln \left( \frac{m^{2}}{\mu^{2}} \right) - 2 J_{s} \left( \frac{p^{2}}{m^{2}} \right) \right] \,.
}
The $1/\epsilon$ pole is canceled by ${\cal M}^{\rm c.t.}$.
We again note that while $Z_{M^{2}}$ is pre-determined, $Z_{\eta \phi}$ is determined to cancel this pole:
\eqs{
\label{eq:two-scalar-Zetaphi}
Z_{\eta \phi} - 1 = \frac{\chi}{16 \pi^{2}} \frac{\eta_{\psi}}{\eta_{\phi}} \frac{1}{\epsilon} \,.
}
From \cref{eq:two-scalar-beta,eq:two-scalar-Zphi}, one obtains
\eqs{
\label{eq:two-scalar-beta_eta}
\frac{d \eta_{\phi}}{d \ln \mu} = \frac{\chi}{16 \pi^{2}} \eta_{\psi} \,.
}
The $m^{2} \ln (m^{2} / \mu^{2})$ term in ${\cal M}^{\rm loop}_{1}$ cancels that in ${\cal M}^{\rm tree}$. 
The $J_{s}$ term is proportional to $p^{2}$ in the heavy limit of $\psi$.
Here
\eqs{
J_{s} (r) &=  \frac{1}{2} \int^{1}_{0} dx \ln \left[ x (x-1) r + 1 - i \epsilon_{\rm ad} \right] \\
&=
\begin{cases}
	 - 1 + \sqrt{\dfrac{4 - r}{r}} \arcsin \dfrac{\sqrt{r}}{2} & \text{(for  $r < 4$)} \\[6pt]
	 - 1 + \sqrt{\dfrac{r - 4}{r}} \left[{\rm arccosh \,} \dfrac{\sqrt{r}}{2} - i \dfrac{\pi}{2} \right] & \text{(for $r > 4$)}
\end{cases} \,.
\label{eq:Jsfunc}
}
For $r > 4$, one needs to take into account an adiabatic parameter $\epsilon_{\rm ad} > 0$ properly.%
\footnote{
Note that an adiabatic parameter $\epsilon_{\rm ad}$ associated with a Wick rotation is different from $\epsilon = 4 - d$ for dimensional regularization.
}
This arises from the fact that the loop scalar can be real.
We remark that $J_{s} (r) \to - r / 12$ ($r \to 0$) in the heavy $\psi$ limit ($p^{2} \ll m_{\rm phys}^{2}$).

The resultant $m^{2}$ term in ${\cal M}^{\rm tree}$ is canceled by the trace-anomaly contribution from the insertion of $T^{\mu}_{~ \mu} \supset (1/4) \epsilon {\tilde \mu}^{\epsilon} \chi \phi^{2} \psi^{2}$.
This contribution does not vanish in the limit of $\epsilon \to 0$ and actually \cref{fig:two-scalar-loop} (b) gives
\eqs{
\label{eq:two-scalar-anom}
i {\cal M}^{\rm loop}_{2}
& = - i \frac{\chi}{16 \pi^{2}} m^{2} \,.
}
We remark that this contribution can be reproduced by the insertion of $- (1/2) \beta_{M^{2}} \phi^{2}$, where $\beta_{M^{2}}$ is the $\beta$ function of $M^{2}$ in \cref{eq:two-scala-beta_M}.

Summing up the decay amplitude, we obtain
\eqs{
{\cal M} (\sigma \to \phi \phi) = 2 (M_{\rm phys}^{2} - \eta_{\phi} p^{2} ) - \frac{\chi}{16 \pi^{2}} \eta_{\psi} p^{2} \left[ \ln \left( \frac{m^{2}}{\mu^{2}} \right) + 2 J_{s} \left( \frac{p^{2}}{m^{2}} \right)\right] + \frac{\chi}{16 \pi^{2}} m^{2} 2 J_{s} \left( \frac{p^{2}}{m^{2}} \right)  \,.
}
In the heavy mass limit of $\psi$ ($p^{2} \ll m_{\rm phys}^{2}$), the decay amplitude is approximated by
\eqs{
\label{eq:two-scalar-MUV}
{\cal M} (\sigma \to \phi \phi) & = 2 M_{\rm phys}^{2} - 2 \left( \eta_{\phi} + \frac{1}{2} \frac{\chi}{16 \pi^{2}} \eta_{\psi} \ln \left( \frac{m^{2}}{\mu^{2}} \right) + \frac{1}{12} \frac{\chi}{16 \pi^{2}} \right) p^{2} \,.
}
Now the dangerous term proportional to $m^{2}$ is absent and thus heavy degrees of freedom decouple from the low-energy dynamics.

Meanwhile, in the low-energy effective theory with almost free ($\lambda_{\phi} \approx 0$) light $\phi$, the leading contribution comes from
\eqs{
(T_{\rm low})^{\mu}_{~ \mu} & = \eta_{\phi {\rm low}} \partial^{2} \phi^{2} + M_{\rm phys}^{2} \phi^{2} \,.
}
We note that the pole mass squared $M_{\rm phys}^{2}$ in the low-energy effective theory is identical to that in the high-energy theory.
The decay amplitude is
\eqs{
\label{eq:two-scalar-MIR}
{\cal M} (\sigma (p) \to \phi (q) \phi (k)) = 2 M_{\rm phys}^{2} - 2 \eta_{\phi {\rm low}} p^{2} \,.
}
By matching \cref{eq:two-scalar-MUV,eq:two-scalar-MIR}, we find
\eqs{
\label{eq:two-scalar-matching}
\eta_{\phi {\rm low}} = \eta_{\phi} + \frac{1}{2} \frac{\chi}{16 \pi^{2}} \eta_{\psi} \ln \left( \frac{m^{2}}{\mu^{2}} \right) + \frac{1}{12} \frac{\chi}{16 \pi^{2}} \,.
}
The last two terms are threshold corrections.
The second term originates from the difference in the $\beta$ function of $\eta$ between the low-energy effective theory ($d \eta_{\rm low} / d \ln \mu = 0$) and the high-energy theory [see \cref{eq:two-scalar-beta_eta}].
Interestingly, the third term is independent of $\eta$ and discussed further in the next section.

\subsection{Yukawa theory}
Let us consider the following action with a scalar field $\phi$%
\footnote{
We consider a pseudo-real scalar field.
We should take into account a tadpole diagram for real scalar fields, while we do not have to do for pseudo-real scalar fields.
}
and a singlet fermion $\psi$:
\eqs{
S_{\rm mat} =& \int d^{d} x \sqrt{- g} \left( \frac{1}{2}  g^{\mu \nu} \nabla_{\mu} \phi_{0} \nabla_{\nu} \phi_{0} + \frac{1}{2} \xi_{0} R \phi_{0}^{2} - \frac{1}{2} M^{2}_{0} \phi_{0}^{2} - \frac{1}{4!} \lambda_{0} \phi_{0}^{4}  \right. \\
& \left. + \frac{1}{2} \overline{\psi}_{0} \gamma^{\mu} i D_{\mu} \psi_{0} - \frac{1}{2} i D_{\mu} \overline{\psi}_{0} \gamma^{\mu} \psi_{0} - m_{0} \overline{\psi}_{0} \psi_{0} - i y_{0} \phi_{0} \overline{\psi}_{0} \gamma_{5} \psi_{0} \right) \,,
}
with $D_{\mu}$ being the local Lorentz and diffeomorphism covariant derivative.
Parameters are a scalar mass $M$, a fermion mass $m$, a self-coupling $\lambda$, and a Yukawa coupling $y$.
We summarize the multiplicative renormalization of fields and parameters and its one-loop expressions in \cref{sec:Yukawa}.
The $d$-dimensional flat-spacetime energy-momentum tensor is given by
\eqs{
T_{\mu \nu} =& \partial_{\mu} \phi_{0} \partial_{\nu} \phi_{0} - \frac{i}{4} \left( \partial_{\{ \mu} \overline{\psi}_{0} \gamma_{\nu \}} \psi_{0} - \overline{\psi}_{0} \partial_{\{ \mu} \gamma_{\nu \}} \psi_{0} \right) \\
& - \left( \frac{d - 2}{4 (d - 1)} + \frac{\eta_{0}}{d-1} \right) (\partial_{\mu} \partial_{\nu} - g_{\mu \nu} \partial^{2}) \phi_{0}^{2} - g_{\mu \nu} {\cal L} \,,
}
where the flat-spacetime Lagrangian density ${\cal L}$ is given by \cref{eq:Yukawa-L}.
Here we use $\xi = (d - 2) / (4 (d - 1)) + \eta / (d - 1)$.
Braces denote the symmetrization of Lorentz indices.
The trace of the energy-momentum tensor is
\eqs{
T^{\mu}_{~ \mu} = \eta_{0} \partial^{2} \phi_{0}^{2} + M_{0}^{2} \phi_{0}^{2} + m_{0} \overline{\psi}_{0} \psi_{0} + \epsilon \frac{1}{4!} \lambda_{0} \phi_{0}^{4}
+ \frac{\epsilon}{2} i y_{0} \phi_{0} \overline{\psi}_{0} \gamma_{5} \psi_{0} + {\rm (e.o.m.)} \,,
}
where the last term is proportional to the equation of motion:
\eqs{
{\rm (e.o.m.)} =& \left( 1 - \frac{\epsilon}{2} \right) \phi_{0} \left[ \partial^{2} \phi_{0} + M^{2} \phi_{0} + \frac{4}{4!} \lambda_{0} \phi_{0}^{4} + i y_{0} \overline{\psi}_{0} \gamma_{5} \psi_{0} \right] \\
& + \left( \frac{3}{2} - \frac{\epsilon}{2} \right) \left[ i \slashed{\partial} \overline{\psi}_{0} + m_{0} \overline{\psi}_{0} + i y_{0} \phi_{0} \overline{\psi}_{0} \gamma_{5} \right] \psi_{0} \\
& + \left( \frac{3}{2} - \frac{\epsilon}{2} \right) \overline{\psi}_{0} \left[ - i \slashed{\partial} \psi_{0} + m_{0}  \psi_{0} + i y_{0} \phi_{0} \gamma_{5} \psi_{0} \right] \,.
}

We consider the decay of the scalaron $\sigma$ into two light scalars $\phi$ at the one-loop level.
We assume that $\phi$ is much lighter than $\psi$, $M_{\rm phys} \ll m_{\rm phys}$ (pole mass), and the self-coupling $\lambda$ is negligible.
The leading contributions originate from
\eqs{
T^{\mu}_{~ \mu} \supset& \eta \partial^{2} \phi^{2} + M^{2} \phi^{2} + m \overline{\psi} \psi  + \frac{\epsilon}{2} i {\tilde \mu}^{\epsilon / 2} y \phi \overline{\psi} \gamma_{5} \psi + (Z_{M^{2}} - 1) M^{2} \phi^{2} \,.
}
Here we use the renormalized fields and parameters in \cref{eq:Yukawa-field-renom,eq:Yukawa-param-renom,eq:Yukawa-eta-renom}.
The one-loop decay amplitude is given by
\eqs{
i {\cal M} (\sigma (p) \to \phi(q) \phi(k)) = i {\cal M}^{\rm tree} + i {\cal M}^{\rm loop} + i {\cal M}^{\rm c.t.} \,,
}
where $p$, $q$, and $k$ are external momenta.
Here, the counterterm contribution from $T^{\mu}_{~ \mu} \supset  (Z_{M^{2}} - 1) M^{2} \phi^{2}$ is given by
\eqs{
i {\cal M}^{\rm c.t.} = 2 i (Z_{M^{2}} - 1) M^{2} = - 16 i \frac{y^{2}}{16 \pi^{2}} m^{2} \frac{1}{\epsilon} \,.
}
In the second equality, we use $Z_{M^{2}}$, which is determined to absorb the $1/\epsilon$ pole in the one-loop self-energy of $\phi$ and is given by \cref{eq:Yukawa-Zphi}.
In contrast to the two-scalar theory, we do not include the counterterm for the non-minimal coupling since no $p^{2}/\epsilon$ term appears at this order: $Z_{\eta} = 1$.
From \cref{eq:Yukawa-Zphi,eq:Yukawa-beta}, one obtains
\eqs{
\label{eq:Yukawa-beta_eta}
\frac{d \eta}{d \ln \mu} = 4 \frac{y^{2}}{16 \pi^{2}} \eta \,.
}

The tree-level contribution from $T^{\mu}_{~ \mu} \supset \eta \partial^{2} \phi^{2} + M^{2} \phi^{2}$ is given by
\eqs{
i {\cal M}^{\rm tree} &= 2 i ( M^{2} - \eta p^{2} ) \times Z^{\rm pole}_{\phi} \\
&=  2 i ( M_{\rm phys}^{2} - \eta p^{2} ) + 8 i \frac{y^{2}}{16 \pi^{2}} m^{2} \left[ \ln \left( \frac{m^{2}}{\mu^{2}} \right) - 1 \right] - 4 i \eta p^{2} \frac{y^{2}}{16 \pi^{2}} \ln \left( \frac{m^{2}}{\mu^{2}} \right) \,.
}
In the second line, we use $Z^{\rm pole}_{\phi}$ in \cref{eq:Yukawa-Zphi_pole} and the relation between the pole and renormalized masses squared in \cref{eq:Yukawa-delM}.
The $m^{2}$ term in ${\cal M}^{\rm tree}$ diverges as one takes the heavy limit of $\psi$.
This originates from the fact that the scalar mass squared is sensitive to ultraviolet physics and one needs fine-tuning to realize $M_{\rm phys} \ll m_{\rm phys}$.
This $m^{2}$ term should be canceled by other contributions so that the amplitude is insensitive to ultraviolet physics.
We will see it shortly below.

\begin{figure}
	\centering
	\includegraphics[width=1.0\linewidth]{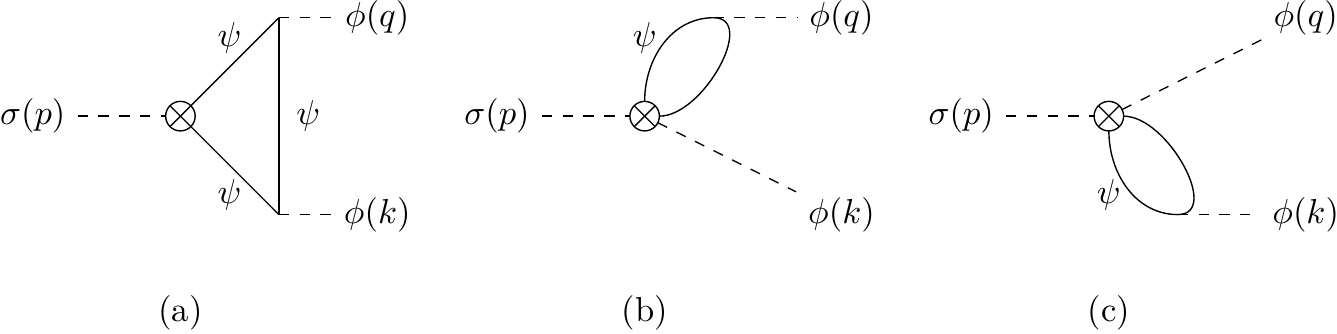}
	\caption{One-loop diagrams for scalaron decay ($\sigma \to \phi \phi$).
	Crossed dots denote the insertion of the energy-momentum tensor, $m \overline{\psi} \psi$ (a) and $(\epsilon / 2) i {\tilde \mu}^{\epsilon / 2} y \phi \overline{\psi} \gamma_{5} \psi$ (b, c).}
	\label{fig:Yukawa-loop}
\end{figure}

As shown in \cref{fig:Yukawa-loop}, several loop diagrams contribute to ${\cal M}^{\rm loop}$.
We provide computational details in \cref{sec:Yukawa}.
\cref{fig:Yukawa-loop} (a) is the one-loop contribution from $T^{\mu}_{~ \mu} \supset m \overline{\psi} \psi$:
\eqs{
i {\cal M}^{\rm loop}_{1} = - 8 i m^{2} \frac{y^{2}}{16 \pi^{2}} \left[ - \frac{2}{\epsilon} + J_{f} \left( \frac{p^{2}}{m^{2}}, \frac{M^{2}}{m^{2}} \right) + \ln \left( \frac{m^{2}}{\mu^{2}} \right) + 2 J_{s} \left( \frac{p^{2}}{m^{2}} \right) \right] \,.
}
The $1/\epsilon$ pole is canceled by ${\cal M}^{\rm c.t.}$.
We again note that $Z_{M^{2}}$ is pre-determined.
The $m^{2} \ln (m^{2} / \mu^{2})$ term cancels that in ${\cal M}^{\rm tree}$.
The $J_{s}$ term is proportional to $p^{2}$ in the heavy limit of $\psi$ [see \cref{eq:Jsfunc}].
We introduce
\eqs{
\label{eq:Jffunc}
J_{f} (r, R) &= \left( \frac{r}{2} - R \right) \int^{1}_{0} dx \int^{x}_{0} dy \frac{1}{- r (1 - x) (x - y) - R \, y (1 - y) + 1 - i \epsilon_{\rm ad}} \,.
}
We remark that $J_{f} (r, R) \to r / 4 - R / 2$ ($r \to 0$, $R \to 0$) in the heavy $\psi$ limit ($p^{2} \ll m_{\rm phys}^{2}$, $M_{\rm phys}^{2} \ll m_{\rm phys}^{2}$). 
The $r$ term is proportional to $p^{2}$, while the $R$ term is proportional to $M^{2}$.
The latter does not diverge, but also does not vanish in the heavy limit of $\psi$.
This $M^{2}$ term should also be can canceled by other contributions so that the amplitude is insensitive to ultraviolet physics.

The resultant $m^{2}$ term in ${\cal M}^{\rm tree}$ and $M^{2}$ term in ${\cal M}^{\rm loop}_{1}$ are canceled by the trace-anomaly contribution from the insertion of $T^{\mu}_{~ \mu} \supset (\epsilon / 2) i {\tilde \mu}^{\epsilon / 2} y \phi \overline{\psi} \gamma_{5} \psi$.
We remark that this contribution does not vanish in the limit of $\epsilon \to 0$ and actually
\cref{fig:Yukawa-loop} (b, c) give
\eqs{
\label{eq:Yukawa-anom}
i {\cal M}^{\rm loop}_{2} = 8 i \frac{y^{2}}{16 \pi^{2}} m^{2} - 4 i \frac{y^{2}}{16 \pi^{2}} M^{2} \,.
}
We remark that this contribution can be reproduced by the insertion of $- (1/2) \beta_{M^{2}} \phi^{2}$, where $\beta_{M^{2}}$ is the $\beta$ function of $M^{2}$ in \cref{eq:Yukawa-beta_M}.

In summary, the total amplitude is given by
\eqs{
{\cal M}(\sigma \to \phi \phi) =& 2 ( M_{\rm phys}^{2} - \eta p^{2} ) - 4 \eta p^{2} \frac{y^{2}}{16 \pi^{2}} \ln \left( \frac{m^{2}}{\mu^{2}} \right) - 4 i \frac{y^{2}}{16 \pi^{2}} M^{2} \\
& - 8 m^{2} \frac{y^{2}}{16 \pi^{2}} \left[ J_{f} \left( \frac{p^{2}}{m^{2}}, \frac{M^{2}}{m^{2}} \right) + 2 J_{s} \left( \frac{p^{2}}{m^{2}} \right) \right] \,.
}
In the heavy $\psi$ limit ($p^{2} \ll m_{\rm phys}^{2}$, $M_{\rm phys}^{2} \ll m_{\rm phys}^{2}$), the total amplitude is approximated by
\eqs{
\label{eq:Yukawa-MUV}
{\cal M}(\sigma \to \phi \phi) = 2 M_{\rm phys}^{2} - 2 \left( \eta + 2 \frac{y^{2}}{16 \pi^{2}} \eta \ln \left( \frac{m^{2}}{\mu^{2}} \right) + \frac{1}{3} \frac{y^{2}}{16 \pi^{2}} \right) p^{2} \,.
}
Now the dangerous term proportional to $m^{2}$ or $M^{2}$ is absent and thus heavy degrees of freedom decouple from the low-energy dynamics.

Meanwhile, in the low-energy effective theory with almost free ($\lambda \approx 0$) light $\phi$, the leading contribution comes from
\eqs{
(T_{\rm low})^{\mu}_{~ \mu} & = \eta_{\rm low} \partial^{2} \phi^{2} + M_{\rm phys}^{2} \phi^{2} \,.
}
We note that the pole mass squared $M_{\rm phys}^{2}$ in the low-energy effective theory is identical to that in the high-energy theory.
The decay amplitude is
\eqs{
\label{eq:Yukawa-MIR}
{\cal M} (\sigma (p) \to \phi (q) \phi (k)) = 2 M_{\rm phys}^{2} - 2 \eta_{\rm low} p^{2} \,.
}
By matching \cref{eq:Yukawa-MUV,eq:Yukawa-MIR}, we find
\eqs{
\label{eq:Yukawa-matching}
\eta_{\rm low} =  \eta + 2 \frac{y^{2}}{16 \pi^{2}} \eta \ln \left( \frac{m^{2}}{\mu^{2}} \right) + \frac{1}{3} \frac{y^{2}}{16 \pi^{2}}  \,.
}
The last two terms are threshold corrections.
The second term originates from the difference in the $\beta$ function of $\eta$ between the low-energy effective theory ($d \eta_{\rm low} / d \ln \mu = 0$) and the high-energy theory [see \cref{eq:Yukawa-beta_eta}].
Interestingly, the third term is independent of $\eta$ and discussed further in the next section.

\section{Quantum-induced value of $\eta$ \label{sec:naturaleta}}

We again note that seemingly we have studied the scalaron decay amplitude in the previous section, but this is just for diagrammatic convenience.
In the following, we regard $\phi$ as the inflaton and the matching condition in \cref{eq:two-scalar-matching,eq:Yukawa-matching} as that for the non-minimal coupling of the inflaton.
Interestingly, a threshold correction that is independent of $\eta$ appears at the one-loop order [see the last terms in \cref{eq:two-scalar-matching,eq:Yukawa-matching}].
This threshold correction is a quantum-induced value of $\eta$ in the low-energy theory because it is induced irrespective of our choice of $\eta$ in the high-energy theory.

A caveat is that this low-energy theory itself may not be ``natural'' in the sense that one has to fine-tune the scalar mass squared to keep $\phi$ light.
This originates from the fact that the quantum correction to the scalar mass squared is quadratic divergent as intensively discussed in the context of the Higgs mass squared in the standard model~\cite{Gildener:1976ai, Weinberg:1978ym, tHooft:1979rat} (see Ref.~\cite{Giudice:2008bi} for a review).
In the two-scalar theory and Yukawa theory, we need fine-tuning between the renormalized mass of $\phi$ and the mass of heavy degrees of freedom [see \cref{eq:two-scalar-delM,eq:Yukawa-delM}].
This is known to be finite naturalness (see Refs.~\cite{Bardeen:1995kv, Shaposhnikov:2007nj} for the standard-model Higgs).
Thus, one may regard this threshold correction as a quantum-induced value in an ``unnatural'' theory.

What is a quantum-induced value of $\eta$ in a ``natural'' theory, i.e., when degrees of freedom that couple to $\phi$ are as light as $\phi$?
It is the {\it inhomogeneous} solution of the RGE of $\eta$:
\eqs{
\label{eq:beta_eta}
\frac{d \eta}{d \ln \mu} = \gamma_{\phi^{2}}^{T} \eta + {\tilde \beta}_{\eta} \,,
}
where $\eta$ and $\gamma_{\phi^{2}}$ should be understood as a vector and matrix, respectively, for multiple scalar fields.
Here
\eqs{
\label{eq:gamma}
& \phi_{0}^{2} = Z_{\phi^{2}} [\phi^{2}]  \,, \\
& \frac{d \ln Z_{\phi^{2}}}{d \ln \mu} = \gamma_{\phi^{2}} \,.
}
The homogeneous term is proportional to the anomalous dimension of $\phi^{2}$ in the RGE.
This is because the renormalization of the scalar field squared is multiplicative, $Z_{\phi^{2}}^{-1}  \partial^{2} \phi_{0}^{2} = \partial^{2} [\phi^{2}]$.
It means that all the counterterms to renormalize $\phi^{2}$ are included in $Z_{\phi^{2}}^{-1} Z_{\rm \phi}$.
${\tilde \beta}_{\eta}$ denotes the inhomogeneous term of the RGE and induces $\Delta \eta$ through the running irrespectively of our initial choice of $\eta$.
This $\Delta \eta$ is nothing but a quantum-induced value of $\eta$.

Before going to the inhomogeneous solution, let us study the homogeneous solution.
We compute $Z_{\phi^{2}}$ in the two-scalar theory in \cref{sec:two-scalar}.
From \cref{eq:two-scalar-beta,eq:two-scalar-Z-com}, one obtains
\eqs{
\gamma_{\phi^{2}} =
\begin{pmatrix}
0 & \dfrac{ \chi}{16 \pi^{2}} \\[6pt]
\dfrac{\chi}{16 \pi^{2}} & 0
\end{pmatrix}
\,,
}
for $\eta = (\eta_{\phi}, \eta_{\psi})^{T}$.
Now one sees that \cref{eq:beta_eta} reproduces \cref{eq:two-scalar-beta_eta}.
By using the $\beta$ function of $\chi$ [see \cref{eq:two-scalar-beta_chi}], we find the homogeneous solution:
\eqs{
\label{eq:two-scalar-eta}
\eta_{\phi} & = \eta_{\phi i} \frac{1}{2} \left[ \left( \frac{\chi}{\chi_{i}} \right)^{1/2} + \left( \frac{\chi}{\chi_{i}} \right)^{-1/2} \right] + \eta_{\psi i} \frac{1}{2} \left[ \left( \frac{\chi}{\chi_{i}} \right)^{1/2} - \left( \frac{\chi}{\chi_{i}} \right)^{-1/2} \right] \,, \\
\eta_\psi & = \eta_{\phi i} \frac{1}{2} \left[ \left( \frac{\chi}{\chi_{i}} \right)^{1/2} - \left( \frac{\chi}{\chi_{i}} \right)^{-1/2} \right] + \eta_{\psi i} \frac{1}{2} \left[ \left( \frac{\chi}{\chi_{i}} \right)^{1/2} + \left( \frac{\chi}{\chi_{i}} \right)^{-1/2} \right] \,,
}
where the subscript $i$ denotes the boundary condition of the RGE, i.e., $\eta_{\phi} = \eta_{\phi i}$ and $\eta_{\psi} = \eta_{\psi i}$ at $\chi = \chi_{i}$.

In the Yukawa theory, from \cref{eq:Yukawa-Zphi,eq:Yukawa-beta} with the notion that $Z_{\phi^{2}} = Z_{\phi}$ at the one-loop level, one obtains
\eqs{
\gamma_{\phi^{2}} = 4 \frac{y^{2}}{16 \pi^{2}} \,.
}
Now one sees that \cref{eq:beta_eta} reproduces \cref{eq:Yukawa-beta_eta}.
By using the $\beta$ function of $y$ [see \cref{eq:Yukawa-beta_y}], we get the homogeneous solution:
\eqs{
\label{eq:Yukawa-eta}
\eta = \eta_{i} \left( \frac{y}{y_{i}} \right)^{4/5} \,,
}
where the subscript $i$ denotes the boundary condition of the RGE, i.e., $\eta = \eta_{i}$ at $y = y_{i}$.
$\eta$ diminishes toward low energy as the theory becomes weakly coupled.
$\eta$ vanishes at the Gaussian fixed point, leaving $T^{\mu}_{~ \mu} = 0$.

Now let us discuss the inhomogeneous solution.
To find the inhomogeneous solution of $\eta$ at the $n$-loop order (leading contribution does not need to be at the one-loop order), one has to determine the inhomogeneous term at the $(n+1)$-loop level.
This is because the right-hand side of \cref{eq:beta_eta} is of the order of $(\beta_{\lambda} / \lambda) \eta$ ($\lambda$ correctively denotes couplings).
The inhomogeneous term arises from the $p^{2} / \epsilon$ pole in the trace-anomaly contribution from the insertion of terms proportional to $\epsilon$ in $T^{\mu}_{~ \mu}$ ($p$: incoming momentum) and two scalars outgoing.
In the $\lambda \phi^{4}$ theory, the trace-anomaly term is $T^{\mu}_{~ \mu} \supset (\epsilon / 4!) \lambda_{0} \phi_{0}^{4}$.
${\tilde \beta}_{\eta}$ arises at the four-loop level ($\lambda^{4}$) and leads to the inhomogeneous solution at the three-loop order~\cite{Collins:1976vm, Brown:1980qq, Hathrell:1981zb}:
\eqs{
\eta = \eta_{i} \left( \frac{\lambda}{\lambda_{i}} \right)^{1/3} - \frac{\lambda^{3}}{864 (4\pi)^{6}} \,,
}
where the subscript $i$ denotes the boundary condition of the RGE, i.e., $\eta = \eta_{i}$ at $\lambda = \lambda_{i}$, again.
The first term is the homogeneous solution, while the second term is the inhomogeneous one.
The trace-anomaly term is $T^{\mu}_{~ \mu} \supset (\epsilon / 4) \chi_{0} \phi_{0}^{2} \psi_{0}^{2}$ in the two-scalar theory, and $T^{\mu}_{~ \mu} \supset (\epsilon / 2) i y_{0} \phi_{0} \overline{\psi}_{0} \gamma_{5} \psi_{0}$ in the Yukawa theory.
We find {\it no} $p^{2} / \epsilon$ pole in one-loop diagrams in either the two-scalar theory or Yukawa theory.
Thus, ${\tilde \beta}_{\eta}$ may arise only at the two-loop level in both theories, which may provide the inhomogeneous solution of $\eta$ at the one-loop order.

\section{Conclusion and remarks \label{sec:concl}}

The energy-momentum tensor $T_{\mu \nu}$ determines the coupling of matter to gravity and provides a valuable site where we can study the properties of a non-minimal coupling of a scalar field to gravity even in the flat spacetime.
We have studied the properties of its trace $T^{\mu}_{~ \mu}$, which is tightly related with scale invariance, particularly motivated by its implications for inflation models.
To be concrete, we have worked it out in the two-scalar theory and Yukawa theory.

The first property that we have stressed is the decoupling of heavy degrees of freedom.
This should be held from the effective field theory point of view, but is not apparent at first sight: heavy degrees of freedom appear in $T^{\mu}_{~ \mu}$ in proportion to their mass, which leaves non-decoupling effects at the loop level.
We have demonstrated that the trace-anomaly contribution from terms proportional to $\epsilon$ in dimensional regularization (and thus vanishes at the classical level) cancels the non-decoupling contribution from mass terms.
The similar conclusion is derived in the gauge trace anomaly~\cite{Kamada:2019pmx}.
This conclusion is relevant, e.g., when one considers $R^{2}$ inflation, where the inflaton (scalaron) couples to $T^{\mu}_{~ \mu}$.
One can safely evaluate, e.g., the decay rate of the scalaron, in the effective field theory.

As a byproduct in demonstrating the decoupling, we have found that a non-minimal coupling, i.e., $\xi = 1 / 6 + \eta / 3$ in four dimensions, receives a one-loop-order threshold correction independent of $\eta$.
This threshold correction is a quantum-induced value of $\eta$ in the low-energy theory in the sense that this value is irrespective of our choice of $\eta$ in the high-energy theory.
On the other hand, one may doubt that this low-energy theory itself is ``unnatural,'' since one has to fine-tune the scalar mass squared to keep a light scalar while taking other degrees of freedom heavy.
The situation can change during inflation.
The field value of the inflaton can make some degrees of freedom, which are light in vacuum, heavy.
Integrating them out leaves a threshold correction to the inflaton.
The threshold correction can even change as the field value of the inflaton changes during inflation, as seen in \cref{eq:two-scalar-matching,eq:Yukawa-matching}, supposing that $m^{2}$ is a function of the field value of the inflaton.
It deserves a further investigation, but we do not go into detail in this article.

To discuss a quantum-induced value of $\eta$ in a ``natural'' theory, we have taken notice of the inhomogeneous solution of the RGE of $\eta$.
The inhomogeneous term of the RGE induces the inhomogeneous solution of $\Delta \eta$ irrespectively of our choice of $\eta$ and thus $\Delta \eta$ is a quantum-induced value of $\eta$.
Since the inhomogeneous term arises from the (composite-operator) renormalization of trace-anomaly terms, $\Delta \eta$ originates from the quantum breaking of scale invariance.

Finally, let us discuss applications to inflation models.
To resurrect chaotic inflation with a single power-law potential, one can introduce $\xi \sim - 10^{-3}$~\cite{Linde:2011nh, Boubekeur:2015xza, Shokri:2019rfi}.
It is very intriguing if this small negative value originates from a quantum-induced value of $\eta$.
To this end, a quantum-induced value should appear at the one-loop order.
The threshold correction found in the two-scalar theory and Yukawa theory is at the one-loop order, but its sign is positive.
Furthermore, we have found no inhomogeneous term at the one-loop level, in either the two-scalar theory or Yukawa theory.
We will study the inhomogeneous term at the two-loop level, which provides an inhomogeneous solution at the one-loop order, in the companion article~\cite{Kamada:2019hpp}.

One has to be careful also in the homogeneous term of $\eta$ when considering inflation models.
$\xi \sim - 10^{-3}$ means that one sets $\eta \simeq - 1 / 2$ at the tree level, and thus $\eta_{i} \simeq - 1 / 2$ in the homogeneous solutions of \cref{eq:two-scalar-eta,eq:Yukawa-eta}.
It can run with the coupling, i.e., $\chi$ in the two-scalar theory and $y$ in the Yukawa theory, during inflation.
The renormalization scale $\mu$ would be set to be the Hubble expansion rate, where the flat-spacetime approximation breaks down, or the mass of loop particles, which is induced by the field value of the inflaton, as for the Coleman-Weinberg potential~\cite{Coleman:1973jx}.
Again it deserves a further investigation, but we do not go into detail in this article.

\subsection*{Acknowledgement}
The work of A. K. and T. K. is supported by IBS under the project code, IBS-R018-D1.
We thank Heejung Kim for discussions in the early stage of this work.
We also thank Kohei Kamada and Kazuya Yonekura for sharing comments on the manuscript.

\bibliographystyle{./utphys}
\bibliography{./ref}

\newpage
\appendix

\section{One-loop calculations \label{sec:oneloop}}

We use the $\overline{\rm MS}$ scheme with the spacetime dimension of $d = 4 - \epsilon$ and the renormalization scale $\mu$.
We compensate a mass dimension by the modified renormalization scale ${\tilde \mu}$, which is defined as
\eqs{
{\tilde \mu}^{2} = \mu^{2} \frac{e^{\gamma_{E}}}{4 \pi}
}
with $\gamma_{E} \simeq 0.577$ being Euler's constant.
One-loop functions are summarized in \cref{sec:loopfunc}.
The arguments of the one-loop functions are omitted when they are obvious.

\subsection{Two-scalar theory \label{sec:two-scalar}}
The Lagrangian density is
\eqs{
\label{eq:two-scalar-L}
{\cal L} = \frac{1}{2} (\partial_{\mu} \phi_{0})^{2} - \frac{1}{2} M^{2}_{0} \phi_{0}^{2} + \frac{1}{2} (\partial_{\mu} \psi_{0})^{2} - \frac{1}{2} m^{2}_{0} \psi_{0}^{2} - \frac{1}{4!} \lambda_{\phi 0} \phi_{0}^{4} - \frac{1}{4!} \lambda_{\psi 0} \psi_{0}^{4} - \frac{1}{4} \chi_{0} \phi_{0}^{2} \psi_{0}^{2} \,.
}
Multiplicative renormalization is set for fields as
\eqs{
\label{eq:two-scalar-field-renom}
\psi_{0} = Z_{\psi}^{1/2} \psi  \,, \quad \phi_{0} = Z_{\phi}^{1/2} \phi \,,
}
and for parameters as
\eqs{
\label{eq:two-scalar-param-renom}
Z_{\phi} M_{0}^{2} = Z_{M^{2}} M^{2} \,, \quad Z_{\psi} m_{0}^{2} = Z_{m^{2}} m^{2} \,, \quad Z_{s}^{2} \lambda_{s 0}  = Z_{\lambda s} {\tilde \mu}^{\epsilon} \lambda_{s} \quad (s = \phi, \psi) \,, \quad Z_{\phi} Z_{\psi} \chi_{0} = Z_{\chi} {\tilde \mu}^{\epsilon} \chi \,,
}
and
\eqs{
\label{eq:two-scalar-eta-renom}
Z_{s} \eta_{s 0} = Z_{\eta s} \eta_{s} \quad (s = \phi, \psi) \,.
}
The Lagrangian density in terms of the renormalized quantities is
\eqs{
{\cal L} = & \frac{1}{2} (\partial_{\mu} \phi)^{2} - \frac{1}{2} M^{2} \phi^{2} + \frac{1}{2} (\partial_{\mu} \psi)^{2} - \frac{1}{2} m^{2} \psi^{2} - \frac{1}{4!} {\tilde \mu}^{\epsilon} \lambda_{\phi} \phi^{4} - \frac{1}{4!} {\tilde \mu}^{\epsilon} \lambda_{\psi} \psi^{4} - \frac{1}{4} {\tilde \mu}^{\epsilon} \chi \phi^{2} \psi^{2} \\
& + \frac{1}{2} (Z_{\phi} - 1) (\partial_{\mu} \phi)^{2} - \frac{1}{2} (Z_{M^{2}} - 1) M^{2} \phi^{2} + \frac{1}{2} (Z_{\psi} - 1) (\partial_{\mu} \psi)^{2} - \frac{1}{2} (Z_{m^{2}} - 1) m^{2} \psi^{2} \\
& - \frac{1}{4!} (Z_{\lambda \phi} - 1) {\tilde \mu}^{\epsilon} \lambda_{\phi} \phi^{4} - \frac{1}{4!} (Z_{\lambda \psi} - 1) {\tilde \mu}^{\epsilon} \lambda_{\psi} \psi^{4} - \frac{1}{4} (Z_{\chi} - 1) {\tilde \mu}^{\epsilon} \chi \phi^{2} \psi^{2} \,.
}
It follows that
\eqs{
\label{eq:two-scalar-beta}
& \beta^{\epsilon}_{\lambda s} = \lambda_{s} \left(  - \epsilon + 2 \frac{d \ln Z_{s}}{d \ln \mu} - \frac{d \ln Z_{\lambda s}}{d \ln \mu} \right) \quad (s = \phi, \psi) \,, \\
& \beta^{\epsilon}_{\chi} = \chi \left( - \epsilon + \frac{d \ln Z_{\phi}}{d \ln \mu} + \frac{d \ln Z_{\psi}}{d \ln \mu} - \frac{d \ln Z_{\chi}}{d \ln \mu} \right) \,, \\
& \beta_{M^{2}} = M^{2} \left( \frac{d \ln Z_{\phi}}{d \ln \mu} - \frac{d \ln Z_{M^{2}}}{d \ln \mu} \right) \,, \\
& \beta_{m^{2}} = m^{2} \left( \frac{d \ln Z_{\psi}}{d \ln \mu} - \frac{d \ln Z_{m^{2}}}{d \ln \mu} \right) \,, \\
& \beta_{\eta s} = \eta_{s} \left( \frac{d \ln Z_{s}}{d \ln \mu} - \frac{d \ln Z_{\eta s}}{d \ln \mu} \right) \quad (s = \phi, \psi) \,.
}

The one-loop self-energy of $\phi$ with the momentum $p$ is given by $i \Pi^{\rm loop} + i \Pi^{\rm c.t.}$ with
\eqs{
& i \Pi^{\rm loop} = \frac{1}{2} (- i {\tilde \mu}^{\epsilon} \chi) \int \frac{d^{d} \ell}{(2 \pi)^{d}} \frac{i}{\ell^{2} - m^{2}} = \frac{i \chi}{16 \pi^{2}} \frac{1}{2} A (m^{2}) \,,\\
& i \Pi^{\rm c.t.} = i (Z_{\phi} - 1) p^{2} - i (Z_{M^{2}} - 1) M^{2} \,.
}
$Z_{\phi}$ and $Z_{M^{2}}$ are determined to cancel $1 / \epsilon$ poles as
\eqs{
\label{eq:two-scalar-Zphi}
Z_{\phi} - 1 = 0 \,, \quad Z_{M^{2}} - 1 = \frac{\chi}{16 \pi^{2}} \frac{m^{2}}{M^{2}} \frac{1}{\epsilon} \,.
}
From \cref{eq:two-scalar-beta}, one finds
\eqs{
\label{eq:two-scala-beta_M}
\beta_{M^{2}} = \frac{\chi}{16 \pi^{2}} m^{2} \,.
}
The resultant self-energy of $\phi$ is
\eqs{
\Gamma_{2} (p^{2})
& = p^{2} - M^{2} - \frac{\chi}{16 \pi^{2}} \frac{1}{2} m^{2} \left[  \ln \left( \frac{m^{2}}{\mu^{2}} \right) - 1 \right] \,.
}
The pole mass squared of $\phi$ satisfies $\Gamma_{2} (M_{\rm phys}^{2}) = 0$.
The difference between the pole and $\overline{\rm MS}$ masses squared is
\eqs{
\label{eq:two-scalar-delM}
M_{\rm phys}^{2} = M^{2} (\mu) + \frac{\chi}{16 \pi^{2}} \frac{1}{2} m^{2} \left[ \ln \left( \frac{m^{2}}{\mu^{2}} \right) - 1 \right] \,.
}
Similarly, for the one-loop self-energy of $\psi$,
\eqs{
\label{eq:two-scalar-Zpsi}
Z_{\psi} - 1 = 0 \,, \quad Z_{m^{2}} - 1 = \frac{\chi}{16 \pi^{2}} \frac{M^{2}}{m^{2}} \frac{1}{\epsilon} \,.
}

\begin{figure}
	\centering
	\includegraphics[width=0.2\linewidth]{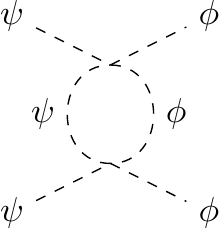}
	\caption{One-loop diagram for the $\phi^{2} \psi^{2}$ interaction.}
	\label{fig:two-scalar-ren_chi}
\end{figure}

We compute the $\beta$ function of $\chi$.
\cref{fig:two-scalar-ren_chi} with the momentum exchange $p$ gives
\eqs{
i {\cal M}^{\rm loop} & = (- i {\tilde \mu}^{\epsilon} \chi)^{2}  \int \frac{d^{d} \ell}{(2 \pi)^{d}} \frac{i}{\ell^{2} - m^{2}} \frac{i}{(\ell + p)^{2} - M^{2}} \\
& = i \frac{\chi^{2}}{16 \pi^{2}} {\tilde \mu}^{\epsilon} B_{0} (p^{2}; m^{2}, M^{2}) \,.
}
Its divergent part is canceled by the counterterm,
\eqs{
i {\cal M}^{\rm c.t.} = - i (Z_{\chi} - i) {\tilde \mu}^{\epsilon} \chi \,,
}
as
\eqs{
Z_{\chi} - 1 = 2 \frac{\chi}{16 \pi^{2}} \frac{1}{\epsilon} \,.
}
From \cref{eq:two-scalar-beta,eq:two-scalar-Zphi,eq:two-scalar-Zpsi}, we obtain the one-loop $\beta$ function of $\chi$:
\eqs{
\label{eq:two-scalar-beta_chi}
\beta_{\chi}^{\epsilon} = - \epsilon \chi + 2 \frac{\chi}{16 \pi^{2}} = - \epsilon \chi + \beta_{\chi}   \,.
}

\begin{figure}
	\centering
	\includegraphics[width=0.4\linewidth]{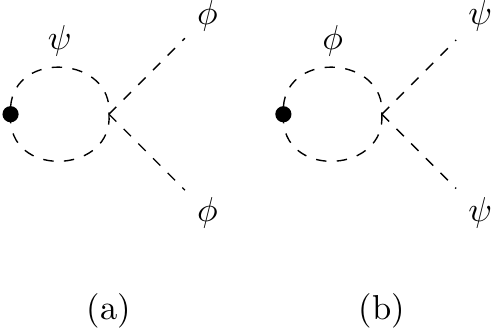}
	\caption{One-loop diagrams for the renormalization of the scalar fields squared.
	Black dots denote the insertion of composite operators, $\psi^{2}$ (a) and $\phi^{2}$ (b).}
	\label{fig:two-scalar-ren_phisq}
\end{figure}

We consider the (composite-operator) renormalization of the scalar fields squared, which is useful in understanding the RGE of the non-minimal couplings:
\eqs{
\begin{pmatrix}
	\phi_{0}^{2} \\
	\psi_{0}^{2}
\end{pmatrix}
= Z_{\phi^{2}}
\begin{pmatrix}
	[\phi^{2}] \\
	[\psi^{2}]
\end{pmatrix} \,,
}
where $Z_{\phi^{2}}$ is understood as a matrix.
In practice, it is convenient to rewrite it as
\eqs{
\label{eq:}
\begin{pmatrix}
	[\phi^{2}] \\
	[\psi^{2}]
\end{pmatrix}
= Z_{\phi^{2}}^{-1}
\begin{pmatrix}
	Z_{\phi} & 0 \\
	0 & Z_{\psi}
\end{pmatrix}
\begin{pmatrix}
	\phi^{2} \\
	\psi^{2}
\end{pmatrix}
=
\begin{pmatrix}
	z_{11} & z_{12} \\
	z_{21} & z_{22}
\end{pmatrix}
\begin{pmatrix}
	\phi^{2} \\
	\psi^{2}
\end{pmatrix} \,.
}
At the tree level, the diagonal components are unity ($z_{11} = z_{22} = 1$), while the off-diagonal components are zero ($z_{12} = z_{21} = 0$).
Since we assume that the self-couplings, $\lambda_{s}$ ($s = \phi, \psi$), are negligible and focus on the quartic coupling $\chi$, no divergence appears in $z_{11}$ and $z_{22}$ and thus at the one-loop level, and thus $z_{11} = z_{22} = 1$.

The one-loop amplitude with the insertion of $[\psi]^{2} \supset z_{21} \phi^{2} + \psi^{2}$, the incoming momentum $p$, and two $\phi$'s outgoing is $i {\cal M} = i {\cal M}^{\rm loop} + i {\cal M}^{\rm c.t.}$ with
\eqs{
i {\cal M}^{\rm c.t.}(\phi^{2} \to \phi \phi) = 2 i z_{21} \,.
}
\cref{fig:two-scalar-ren_phisq} (a) gives
\eqs{
i {\cal M}^{\rm loop}(\psi^{2} \to \phi \phi) & = (- i {\tilde \mu}^{\epsilon} \chi) \int \frac{d^{d} \ell}{(2 \pi)^{d}} \frac{i}{\ell^{2} - m^{2}} \frac{i}{(\ell + p)^{2} - m^{2}} \\
& = - i \frac{\chi}{16 \pi^{2}} B_{0} (p^{2}; m^{2}, m^{2}) \,.
}
$z_{21}$ is determined to cancel the divergence as
\eqs{
z_{21} = \frac{\chi}{16 \pi^{2}} \frac{1}{\epsilon} \,.
}

Similarly, for the one-loop amplitude with the insertion of $[\phi]^{2} \supset \phi^{2} + z_{12} \psi^{2}$, one obtains
\eqs{
i {\cal M}^{\rm c.t.}(\psi^{2} \to \psi \psi) = 2 i z_{12} \,,
}
and
\eqs{
i {\cal M}^{\rm loop}(\phi^{2} \to \psi \psi) & = (- i {\tilde \mu}^{\epsilon} \chi) \int \frac{d^{d} \ell}{(2 \pi)^{d}} \frac{i}{\ell^{2} - M^{2}} \frac{i}{(\ell + p)^{2} - M^{2}} \\
& = - i \frac{\chi}{16 \pi^{2}} B_{0} (p^{2}; M^{2}, M^{2})\,,
}
from \cref{fig:two-scalar-ren_phisq} (b).
$z_{21}$ is determined to cancel the divergence as
\eqs{
z_{12} = \frac{\chi}{16 \pi^{2}} \frac{1}{\epsilon} \,.
}

Since $Z_{\phi} = Z_{\psi} = 1$ at the one-loop level [see \cref{eq:two-scalar-Zphi,eq:two-scalar-Zpsi}],
\eqs{
\label{eq:two-scalar-Z-com}
Z_{\phi^{2}}^{-1}
=
\begin{pmatrix}
	1 & \dfrac{\chi}{16 \pi^{2}} \dfrac{1}{\epsilon} \\[6pt]
	\dfrac{\chi}{16 \pi^{2}} \dfrac{1}{\epsilon} & 1
\end{pmatrix} \,, \quad
Z_{\phi^{2}}
=
\begin{pmatrix}
	1 & - \dfrac{\chi}{16 \pi^{2}} \dfrac{1}{\epsilon} \\[6pt]
	- \dfrac{\chi}{16 \pi^{2}} \dfrac{1}{\epsilon} & 1
\end{pmatrix} \,.
}

We compute the one-loop diagrams with the insertion of $T^{\mu}_{~ \mu}$, which appear in the main text.
\cref{fig:two-scalar-loop} (a) gives
\eqs{
i {\cal M}^{\rm loop}_{1} & = (2 m^{2} - 2 \eta_{\psi} p^{2}) (- i {\tilde \mu}^{\epsilon} \chi) \frac{1}{2} \int \frac{d^{d} \ell}{(2\pi)^{d}} \frac{i}{\ell^{2}-m^{2}} \frac{i}{(\ell + p)^{2}-m^{2}} \\
& = - i \frac{\chi}{16 \pi^{2}} (m^{2} - \eta_{\psi} p^{2}) B_{0} (m^{2}) \\
& = - i \frac{\chi}{16 \pi^{2}} (m^{2} - \eta_{\psi} p^{2}) \left[ \frac{2}{\epsilon} - \ln \left(\frac{m^{2}}{\mu^{2}}\right) - 2 J_{s} \left( \frac{p^{2}}{m^{2}} \right)\right] \,,
}
while \cref{fig:two-scalar-loop} (b) gives
\eqs{
i {\cal M}^{\rm loop}_{2} = (i \epsilon {\tilde \mu}^{\epsilon} \chi) \frac{1}{2} \int \frac{d^{d} \ell}{(2\pi)^{d}} \frac{i}{\ell^{2}-m^{2}} = - i \frac{\chi}{16 \pi^{2}} \epsilon \frac{1}{2} A(m^{2}) = - \frac{i}{16 \pi^{2}} \chi m^{2} \,.
}

\subsection{Yukawa theory \label{sec:Yukawa}}
The Lagrangian density is
\eqs{
\label{eq:Yukawa-L}
{\cal L} = \frac{1}{2} \overline{\psi}_{0} i \gamma^{\mu} \partial_{\mu} \psi_{0} - \frac{1}{2} i \partial_{\mu}  \overline{\psi}_{0} \gamma^{\mu} \psi_{0} - m_{0} \overline{\psi}_{0} \psi_{0} + \frac{1}{2} (\partial_{\mu} \phi_{0})^{2} - \frac{1}{2} M_{0}^{2} \phi_{0}^{2} - \frac{1}{4!} \lambda_{0} \phi_{0}^{4} - i y_{0} \phi_{0} \overline{\psi}_{0} \gamma_{5} \psi_{0} \,.
}
Multiplicative renormalization is set for fields as
\eqs{
\label{eq:Yukawa-field-renom}
\psi_{0} = Z_{\psi}^{1/2} \psi \,, \quad \phi_{0} = Z_{\phi}^{1/2} \phi \,.
}
and for parameters as
\eqs{
\label{eq:Yukawa-param-renom}
Z_{\phi} M_{0}^{2} = Z_{M^{2}} M^{2} \,, \quad Z_{\psi} m_{0} = Z_{m} m \,, \quad Z_{\phi}^{2} \lambda_{0}  = Z_{\lambda} {\tilde \mu}^{\epsilon} \lambda \,, \quad Z_{\phi}^{1/2} Z_{\psi} y_{0}  = Z_{y} {\tilde \mu}^{\epsilon / 2} y  \,,
}
and
\eqs{
\label{eq:Yukawa-eta-renom}
Z_{\phi} \eta_{0} = Z_{\eta} \eta \,.
}
The Lagrangian density in terms of the renormalized quantities is
\eqs{
{\cal L} = &
\frac{1}{2} \overline{\psi} i \gamma^{\mu} \partial_{\mu} \psi - \frac{1}{2} i \partial_{\mu}  \overline{\psi} \gamma^{\mu} \partial_{\mu} \psi - m \overline{\psi} \psi + \frac{1}{2} (\partial_{\mu} \phi)^{2} - \frac{1}{2} M^{2} \phi^{2} - \frac{1}{4!} {\tilde \mu}^{\epsilon} \lambda \phi^{4} - i  {\tilde \mu}^{\epsilon / 2} y \phi \overline{\psi} \gamma_{5} \psi \\
& + \frac{1}{2} (Z_{\psi} - 1) \overline{\psi} i \gamma^{\mu} \partial_{\mu} \psi - \frac{1}{2} (Z_{\psi} - 1) i \partial_{\mu} \overline{\psi} \gamma^{\mu} \psi - (Z_{m} - 1) m \overline{\psi} \psi + \frac{1}{2} (Z_{\phi} - 1) (\partial_\mu \phi)^{2} - \frac{1}{2} (Z_{M^{2}} - 1) M^{2} \phi^{2} \\
& - \frac{1}{4!} (Z_{\lambda} - 1) {\tilde \mu}^{\epsilon} \lambda \phi^{4} - i (Z_{y} - 1) {\tilde \mu}^{\epsilon / 2} y \phi \overline{\psi} \gamma_{5} \psi \,,
}
It follows that
\eqs{
\label{eq:Yukawa-beta}
& \beta^{\epsilon}_{y} = y \left( - \frac{1}{2} \epsilon + \frac{1}{2} \frac{d \ln Z_{\phi}}{d \ln \mu} + \frac{d \ln Z_{\psi}}{d \ln \mu} - \frac{d \ln Z_{y}}{d \ln \mu} \right) \,, \\
& \beta^{\epsilon}_{\lambda} = \lambda \left( - \epsilon + 2 \frac{d \ln Z_{\phi}}{d \ln \mu} - \frac{d \ln Z_{\lambda}}{d \ln \mu} \right) \,, \\
& \beta_{M^{2}} = M^{2} \left( \frac{d \ln Z_{\phi}}{d \ln \mu} - \frac{d \ln Z_{M^{2}}}{d \ln \mu} \right) \,, \\
& \beta_{m^{2}} = m^{2} \left( \frac{d \ln Z_{\psi}}{d \ln \mu} - \frac{d \ln Z_{m^{2}}}{d \ln \mu} \right) \,, \\
& \beta_{\eta} = \eta \left( \frac{d \ln Z_{\phi}}{d \ln \mu} - \frac{d \ln Z_{\eta}}{d \ln \mu} \right) \,.
}

The one-loop self-energy of $\phi$ with the momentum $p$ is given by $i \Pi^{\rm loop} + i \Pi^{\rm c.t.}$ with
\eqs{
i \Pi^{\rm c.t.} & = i (Z_{\phi} - 1)p^{2} - i (Z_{M^{2}} - 1) M^{2} \,, \\
i \Pi^{\rm loop} & = ({\tilde \mu}^{\epsilon/2} y)^{2} (-1) \int \frac{d^{d} \ell}{(2 \pi)^{d}} {\rm tr}\left[ \frac{i (\slashed{\ell} + m)}{\ell^{2} - m^{2}} \gamma_{5}  \frac{i (\slashed{\ell} + \slashed{p} + m)}{(\ell + p)^{2} - m^{2}} \gamma_{5} \right] \\
& = - 4 i \frac{y^{2}}{16 \pi^{2}} \left( A (m^{2}) + p^{\mu}  B_{\mu} (p^{2}; m^{2}, m^{2}) \right)  \\
& = 4 i \frac{y^{2}}{16 \pi^{2}} \left( \frac{1}{2} p^{2}  B_{0} - A \right) \,.
}
Here we use ${\rm tr}[(\slashed{\ell} + m) \gamma_{5} (\slashed{\ell} + \slashed{p} + m) \gamma_{5}] = - 4 (\ell^{2} - m^{2}) - 4 \ell \cdot p$.
The divergent part is
\eqs{
\left( \Pi^{\rm loop} \right)^{\rm pole}_{{\rm of} \, \epsilon} = 8 \frac{y^{2}}{16 \pi^{2}} \left(\frac{1}{2} p^{2} - m^{2} \right) \frac{1}{\epsilon} \,.
}
$Z_{\phi}$ and $Z_{M^{2}}$ are determined to absorb the divergent part as
\eqs{
\label{eq:Yukawa-Zphi}
Z_{\phi} - 1 = - 4 \frac{y^{2}}{16 \pi^{2}} \frac{1}{\epsilon} \,, \quad Z_{M^{2}} - 1 = - 8 \frac{y^{2}}{16 \pi^{2}} \frac{m^{2}}{M^{2}} \frac{1}{\epsilon} \,.
}
From \cref{eq:Yukawa-beta}, one finds
\eqs{
\label{eq:Yukawa-beta_M}
\beta_{M^{2}} = - 8 \frac{y^{2}}{16 \pi^{2}} m^{2} + 4 \frac{y^{2}}{16 \pi^{2}} M^{2} \,.
}
The resultant self-energy of $\phi$ is
\eqs{
\Gamma_{2} (p^{2}) 
& = p^{2} - M^{2} + 4 \frac{y^{2}}{16 \pi^{2}} \left[ \frac{p^{2}}{2} \left[ -  \ln \left( \frac{m^{2}}{\mu^{2}} \right) - 2 J_{s} \left( \frac{p^{2}}{m^{2}} \right) \right] + m^{2} \ln \frac{m^{2}}{\mu^{2}} - m^{2}  \right] \,,
}
where $J_{s} (r)$ is given by \cref{eq:Jsfunc}.
In the heavy mass limit of $\psi$ ($p^{2} \ll m_{\rm phys}^{2}$), $J_{s} (r) \to - r / 12$ ($r \to 0$).
The pole mass squared of the scalar field $\phi$ satisfies $\Gamma_{2} (M_{\rm phys}^{2}) = 0$.
The relation between the pole and $\overline{\rm MS}$ masses squared is
\eqs{
\label{eq:Yukawa-delM}
M_{\rm phys}^{2} = M^{2} (\mu) - 4 \frac{y^{2}}{16 \pi^{2}} m^{2} \left[ \ln \left( \frac{m^{2}}{\mu^{2}} \right) - 1 \right] + 2 \frac{y^{2}}{16 \pi^{2}} M^{2} \ln \left( \frac{m^{2}}{\mu^{2}} \right) \,.
}
The pole wave function is given by
\eqs{
\label{eq:Yukawa-Zphi_pole}
Z^{\rm pole}_{\phi}= \left( \left. \frac{\partial}{\partial p^{2}} \Gamma_{2} \right|_{p^{2} = M_{\rm phys}^{2}} \right)^{-1} = 1 + 2 \frac{y^{2}}{16 \pi^{2}} \ln \left( \frac{m^{2}}{\mu^{2}} \right) \,.
}

The one-loop self-energy of $\psi$ with the momentum $p$ is given by $i \Pi^{\rm loop} + i \Pi^{\rm c.t.}$ with
\eqs{
i \Pi^{\rm loop} & = ({\tilde \mu}^{\epsilon/2} y)^{2} \int \frac{d^{d} \ell}{(2 \pi)^{d}} \gamma_{5} \frac{i (\slashed{\ell} + m)}{\ell^{2} - m^{2}} \gamma_{5} \frac{i}{(\ell - p)^{2} - M^{2}} \\
& = - i \frac{y^{2}}{16 \pi^{2}} \left[ \slashed{p} B_{1}(p^{2}; m^{2}, M^{2}) + m B_{0} \right] \,.
}
The divergent part,
\eqs{
\left( \Pi^{\rm loop} \right)^{\rm pole}_{\rm of \, \epsilon} = \frac{y^{2}}{16 \pi^{2}} \left( \slashed{p} - 2 m \right) \frac{1}{\epsilon} \,,
}
is canceled by the counterterm,
\eqs{
i \Pi^{\rm c.t.} = i ( Z_{\psi} - 1 ) \slashed{p} - i ( Z_{m} - 1 ) m \,.
}
$Z_{\psi}$ and $Z_{m}$ are given by
\eqs{
\label{eq:Yukawa-Zpsi}
Z_{\psi} - 1 = - \frac{y^{2}}{16 \pi^{2}} \frac{1}{\epsilon} \,, \quad Z_{m} - 1 = - 2 \frac{y^{2}}{16 \pi^{2}} \frac{1}{\epsilon} \,.
}

\begin{figure}
	\centering
	\includegraphics[width=0.3\linewidth]{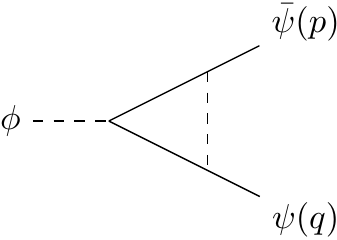}
	\caption{One-loop diagram for the Yukawa interaction.}
	\label{fig:Yukawa-ren_y}
\end{figure}

We compute the $\beta$ function of the Yukawa coupling $y$.
\cref{fig:Yukawa-ren_y} gives
\eqs{
i {\cal M}^{\rm loop} & = ({\tilde \mu}^{\epsilon / 2} y)^{3} \int \frac{d^{d} \ell}{(2 \pi)^{d}}
\gamma_{5} \frac{i (\slashed{\ell} + \slashed{p} + \slashed{q} + m)}{(\ell + p + q)^{2} - m^{2}} \gamma_{5}  \frac{i (\slashed{\ell} + m)}{\ell^{2} - m^{2}} \gamma_{5} \frac{i}{(\ell + q)^{2} -M^{2}} \\
& = ({\tilde \mu}^{\epsilon / 2} y) \frac{y^{2}}{16 \pi^{2}} \gamma_{5} [ - B_{0} (p^{2}; M^{2}, m^{2}) - (\slashed{p} + \slashed{q}) \gamma^{\mu} C_{\mu} (q^{2}, p^{2}, (q + p)^{2}; m^{2}, M^{2}, m^{2}) + (\slashed{p} + \slashed{q}) m C_{0} ] \,.
}
Here we use $(\slashed{\ell} + \slashed{p} + \slashed{q} + m) \gamma_{5} (\slashed{\ell} + m) \gamma_{5} = - (\ell^{2} - m^{2}) + (\slashed{p} + \slashed{q}) (- \slashed{\ell} + m)$.
The divergent part,
\eqs{
\left( i {\cal M}^{\rm loop} \right)^{\rm pole}_{\rm of \, \epsilon} = - 2 ({\tilde \mu}^{\epsilon / 2} y) \frac{y^{2}}{16 \pi^{2}} \frac{1}{\epsilon} \gamma_{5} \,,
}
is canceled by the counterterm,
\eqs{
i {\cal M}^{\rm c.t.} = (Z_{y} - 1) {\tilde \mu}^{\epsilon / 2} y \gamma_{5} \,.
}
$Z_{y}$ is given by
\eqs{
Z_{y} - 1 = 2 \frac{y^{2}}{16 \pi^{2}} \frac{1}{\epsilon} \,.
\label{eq:Yukawa-Zy}
}
From \cref{eq:Yukawa-beta,eq:Yukawa-Zphi,eq:Yukawa-Zpsi}, we get
\eqs{
\label{eq:Yukawa-beta_y}
\beta^{\epsilon}_{y} 
= - \frac{\epsilon}{2} y + \frac{5 y^{3}}{16 \pi^{2}}  = - \frac{\epsilon}{2} y + \beta_{y} \,.
}

We compute the one-loop diagrams with the insertion of $T^{\mu}_{~ \mu}$, which appear in the main text.
\cref{fig:Yukawa-loop} (a) gives
\eqs{
i {\cal M}^{\rm loop}_{1} & =  (i m) ({\tilde \mu}^{\epsilon / 2} y)^{2} (-1) \int \frac{d^{d} \ell}{(2\pi)^{d}} {\rm tr}
\left[ \gamma_{5} \frac{i (\slashed{\ell} + \slashed{p} + m)}{(\ell + p)^{2} - m^{2}} \frac{i (\slashed{\ell} + m)}{\ell^{2} - m^{2}} \gamma_{5} \frac{i (\slashed{\ell} + \slashed{q} + m)}{(\ell + q)^{2} - m^{2}} \right]
+ (q \leftrightarrow k)  \\
& = 4 i m^{2} \frac{y^{2}}{16 \pi^{2}} \left[ B_{0} (p^{2}; m^{2}, m^{2}) + \left( \frac{p^{2}}{2} - M^{2} \right) C_{0} (q^{2}, k^{2}, p^{2}; m^{2}, m^{2}, m^{2}) \right] + (q \leftrightarrow k) \\
& = - 8 i m^{2} \frac{y^{2}}{16 \pi^{2}} \left[ - \frac{2}{\epsilon} + J_{f} \left( \frac{p^{2}}{m^{2}}, \frac{M^{2}}{m^{2}} \right) + \ln \left( \frac{m^{2}}{\mu^{2}} \right) + 2 J_{s} \left( \frac{p^{2}}{m^{2}} \right) \right] \,.
}
Here we use ${\rm tr} [\gamma_{5} (\slashed{\ell} + \slashed{p} + m) (\slashed{\ell} + m) \gamma_{5} (\slashed{\ell} + \slashed{q} + m)] = - 4 m  ((\ell + q)^{2} - m^{2})  - 4 m q \cdot k$.
The $B_{0}$ and $C_{0}$ functions lead to $J_{s}(r)$ in \cref{eq:Jsfunc} and $J_{f} (r, R)$ in \cref{eq:Jffunc}, respectively.

\cref{fig:Yukawa-loop} (b, c) give
\eqs{
i {\cal M}^{\rm loop}_{2} & = \left( - \frac{1}{2} \epsilon {\tilde \mu}^{\epsilon / 2} y \right) ({\tilde \mu}^{\epsilon / 2} y) (-1) \int \frac{d^{d} \ell}{(2\pi)^{d}} {\rm tr} \left[ \frac{i (\slashed{\ell} + m)}{\ell^{2} - m^{2}}\gamma_{5} \frac{ i (\slashed{\ell} + \slashed{q} + m)}{(\ell + q)^{2} - m^{2}} \gamma_{5} \right] + (q \leftrightarrow k) \\
& = 2 i  \frac{y^{2}}{16 \pi^{2}} \epsilon \left[ A (m^{2}) + q^{\mu} B_{\mu} (q^{2}; m^{2}, m^{2}) \right] + (q \leftrightarrow k) \\
& = 8 i \frac{y^{2}}{16 \pi^{2}} m^{2} - 4 i \frac{y^{2}}{16 \pi^{2}} M^{2} \,.
}
Here we use ${\rm tr}[(\slashed{\ell} + m) \gamma_{5} (\slashed{\ell} + \slashed{p} + m) \gamma_{5}] = - 4 (\ell^{2} - m^{2}) - 4 \ell \cdot p$.

\subsection{Summary of one-loop functions \label{sec:loopfunc}}
One-loop functions are based on Refs.~\cite{tHooft:1978jhc, Passarino:1978jh} (see also Appendix F of Ref.~\cite{Logan:1999if}).
The one-point integral is defined as
\eqs{
{\tilde \mu}^{\epsilon} \int \frac{d^{d} \ell}{(2 \pi)^{d}} \frac{1}{\ell^{2} - m^{2}} = \frac{i}{16 \pi^{2}} A (m^{2}) \,.
}
The explicit form is
\eqs{
A = m^{2} \left( \frac{2}{\epsilon} - \ln \left( \frac{m^{2}}{\mu^{2}} \right) + 1 \right) \,.
}

Two-point integrals are defined as
\eqs{
{\tilde \mu}^{\epsilon} \int \frac{d^{d} \ell}{(2 \pi)^{d}} \frac{1; \ell_{\mu}; \ell_{\mu} \ell_{\nu}}{[\ell^{2} - m_{1}^{2}] [(\ell+p)^{2} - m_{2}^{2}]} = \frac{i}{16 \pi^{2}} B_{0; \mu; \mu \nu} (p^{2}; m_{1}^{2}, m_{2}^{2}) \,,
}
where
\eqs{
& B_{\mu} = p_{\mu} B_{1} \,, \\
& B_{\mu \nu} = g_{\mu \nu} B_{22} + p_{\mu} p_{\nu} B_{21} \,.
}
The divergent parts are
\eqs{
& (B_{0})^{\rm pole}_{{\rm of} \, \epsilon} = \frac{2}{\epsilon} \,, \\
& (B_{1})^{\rm pole}_{{\rm of} \, \epsilon} = - \frac{1}{\epsilon} \,, \\
& (B_{22})^{\rm pole}_{{\rm of} \, \epsilon} = \frac{1}{2} \left( m_{1}^{2} + m_{2}^{2} - \frac{p^{2}}{3} \right) \frac{1}{\epsilon} \,, \\
& (B_{21})^{\rm pole}_{{\rm of} \, \epsilon} = \frac{2}{3} \frac{1}{\epsilon} \,.
}

For our purpose, we can take $m_{1} = m_{2} = m$:
\eqs{
& B_{1} = - \frac{1}{2} B_{0} \,, \\
& B_{22} = \frac{1}{6} \left[ A + 2 m^{2} B_{0} - \frac{p^{2}}{2} B_{0} + 2 m^{2} - \frac{p^{2}}{3} \right] \,, \\
& B_{21} = \frac{1}{3 k^{2}} \left[ A - m^{2} B_{0} + p^{2} B_{0} - m^{2} + \frac{p^{2}}{6} \right] \,.
}
The explicit form with a Feynman parameter integral is
\eqs{
B_{0} = \frac{2}{\epsilon} - \int^{1}_{0} dx \ln \left( \frac{m^{2} - x (1 - x) p^{2} - i \epsilon_{\rm ad}}{\mu^{2}} \right) \,.
}

Three-point integrals are defined as
\eqs{
{\tilde \mu}^{\epsilon} \int \frac{d^{d} \ell}{(2 \pi)^{d}} \frac{1; \ell_{\mu}; \ell_{\mu} \ell_{\nu}}{[\ell^{2} - m_{1}^{2}] [(\ell+q)^{2} - m_{2}^{2}] [(\ell+q+k)^{2} - m_{3}^{2}]} = \frac{i}{16 \pi^{2}} C_{0; \mu; \mu \nu} (q^{2}, k^{2}, p^{2}; m_{1}^{2}, m_{2}^{2}, m_{3}^{2}) \,,
}
where $p + q + k = 0$ and
\eqs{
& C_{\mu} = q_{\mu} C_{11} + k_{\mu} C_{12} \,, \\
& C_{\mu \nu} = g_{\mu \nu} C_{24} + q_{\mu} q_{\nu} C_{21} + k_{\mu} k_{\nu} C_{22} + \left( q_{\mu} k_{\nu} +  k_{\mu} q_{\nu} \right) C_{23} \,.
}
The divergent part is
\eqs{
& (C_{24})^{\rm pole}_{{\rm of}\, \epsilon} = \frac{1}{2} \frac{1}{\epsilon} \,,
}
while the other $C$'s are finite.

For our purpose, again we can take $m_{1} = m_{2} = m_{3} = m$ and $q^{2} = k^{2} = M^{2}$.
The explicit form with Feynman parameter integrals is
\eqs{
C_{0} = - \int_{0}^{1} dx \int_{0}^{x} dy \, \frac{1}{- p^{2} (1 - x) (x - y) - M^{2} \, y (1 - y) + m^{2} - i \epsilon_{\rm ad}} \,.
}
For $q^{2} = k^{2} = 0$,
\eqs{
& C_{11} = \frac{1}{p^{2}} \left[ B_{0} (q^{2}) - B_{0} (p^{2}) - p^{2} \, C_{0} \right] \,, \\
& C_{12} = \frac{1}{p^{2}} \left[ B_{0} (p^{2}) - B_{0} (k^{2}) \right] \,, \\
& C_{24} = \frac{1}{4} \left[ B_{0} (p^{2}) + 2 m^{2} C_{0} + 1 \right] \,, \\
& C_{21} = - \frac{1}{2 p^{2}} \left[ 3 B_{0} (p^{2}) - 3 B_{0} (p^{2}) - 2 p^{2} \, C_{0} \right] \,, \\
& C_{23} = - \frac{1}{2 p^{2}} \left[ 2 B_{0} (p^{2}) - 2 B_{0} (k^{2})  + 2 m^{2} C_{0} + 1 \right] \,, \\
& C_{22} = - \frac{1}{2 p^{2}} \left[ B_{0} (p^{2}) - B_{0} (k^{2}) \right] \,.
}
Here we use the shorthand notation of $B_{0} (p^{2}) = B_{0} (p^{2}; m^{2}, m^{2})$.

\end{document}